\documentclass{webofc}
\usepackage[varg]{txfonts}  
\begin{document}
\title{Vector boson tagged jets and jet substructure}
%
%

\author{\firstname{Ivan} \lastname{Vitev}\inst{1}\fnsep\thanks{\email{ivitev@lanl.gov}} 
}

\institute{
Los Alamos National Laboratory, Theoretical Division, Mail Stop B283 \\
Los Alamos, NM 87545, USA 
          }

\abstract{%
In these proceedings,  we report on recent results related to  vector boson-tagged jet production in heavy ion collisions and the related modification of jet 
substructure, such as jet shapes and jet momentum sharing distributions.   $Z^0$-tagging and $\gamma$-tagging of jets  provides new opportunities to study parton shower formation and propagation in the quark-gluon plasma and has been argued to provide tight constrains on the energy loss of reconstructed jets. We present theoretical predictions  for isolated photon-tagged and electroweak boson-tagged jet production in Pb+Pb collisions at $\sqrt{s_{NN}} = 5.02$~TeV at the LHC, addressing  the modification of their transverse momentum and transverse momentum imbalance distributions.  Comparison to recent ATLAS and CMS experimental measurements is performed that can shed light on the medium-induced radiative corrections  and energy dissipation due to collisional processes of predominantly quark-initiated jets.  The modification of parton splitting functions in the QGP further implies that the  substructure of jets in heavy ion collisions may differ significantly  from the corresponding substructure in  proton-proton collisions.  Two such observables and the implication of tagging on their evaluation is also discussed. 
}
\maketitle
\section{Introduction}
\label{intro}

In ultrarelativistic collisions of heavy nuclei a new deconfined state of matter, the quark-gluon plasma (QGP), is expected to be formed. 
The attenuation of the production rate of high transverse momentum particles  in nucleus-nucleus (A+A) relative to the binary collision-scaled 
proton-proton (p+p) reactions has been proposed a long time ago as a signature of its formation~\cite{Wang:1991xy}. The quenching of hadrons opposite 
a direct photon was studied~\cite{Wang:1996yh} shortly after -- a precursor of the modern tagged reconstructed jet 
observables~\cite{Srivastava:2002kg,Neufeld:2010fj}.  The production of a vector boson  in association with jets  is a  powerful channel 
to probe the fundamental properties of Quantum Chromodynamics (QCD).  Vector boson-tagged jets are also very well suited to 
studying the  effects of the QGP. The tagging bosons escape the region of the hot dense medium unaffected and can be used to constrain  the energy of the away-side parton shower.  Previous studies of vector boson tagged jet production in heavy ion collisions have   been performed in 
the  framework of perturbative QCD~\cite{Dai:2012am,Neufeld:2012df}. A  Boltzmann transport model~\cite{Wang:2013cia}, an event generator JEWEL~\cite{KunnawalkamElayavalli:2016ttl} and a hybrid strong/weak coupling model~\cite{Casalderrey-Solana:2015vaa} have also 
been used. On the experimental side,  isolated $\gamma$-tagged and $Z^0$-tagged jets results in Pb+Pb collisions at   $\sqrt{s_{NN}}=5.02$ TeV have become available at the LHC from the ATLAS and CMS collaborations~\cite{Sirunyan:2017jic,CMS:2016ynj,ATLAS:2016tor}. These new measurements call for improved theoretical calculations~\cite{Kang:2017xnc} to interpret the experimental data. 

Vector-boson tagged jet cross section modification cannot be understood in isolation. The emergence of  in-medium parton branching, qualitatively different from the one which determines the jet properties in $e^++e^-$, $e^-+p$, and $p+p$ collisions, also implies non-trivial modification of 
jet substructure observables. These include jet shapes, jet fragmentation functions and jet momentum sharing distributions. In these proceedings 
we also report on calculation of two of the above~\cite{Chien:2015hda,Chien:2016led}  in a theoretical framework that uses the same fundamental 
 input  -- the in-medium splitting functions.
 
\section{Vector boson tagged jets}
Let us now turn to the new theoretical  results on V+jet modification at the LHC.

\subsection{Proton collisions}
\label{subsec-pp}
\begin{figure}[h]
\centering
\includegraphics[width=6.5cm,clip]{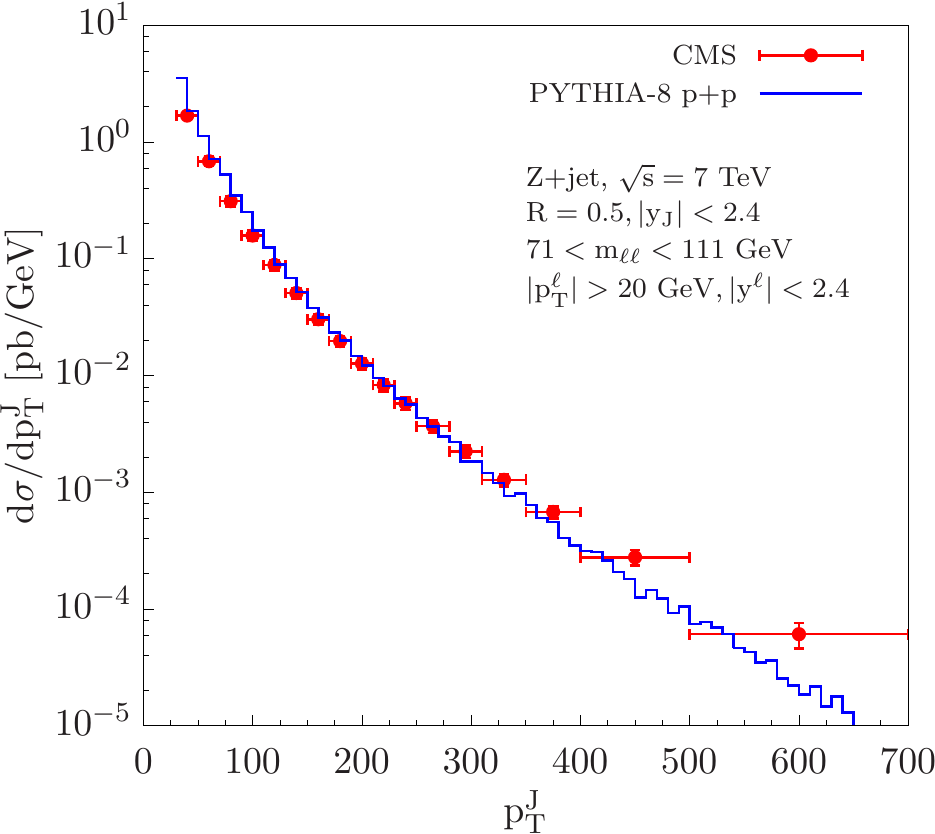} \hspace*{.5cm}
\includegraphics[width=6.5cm,clip]{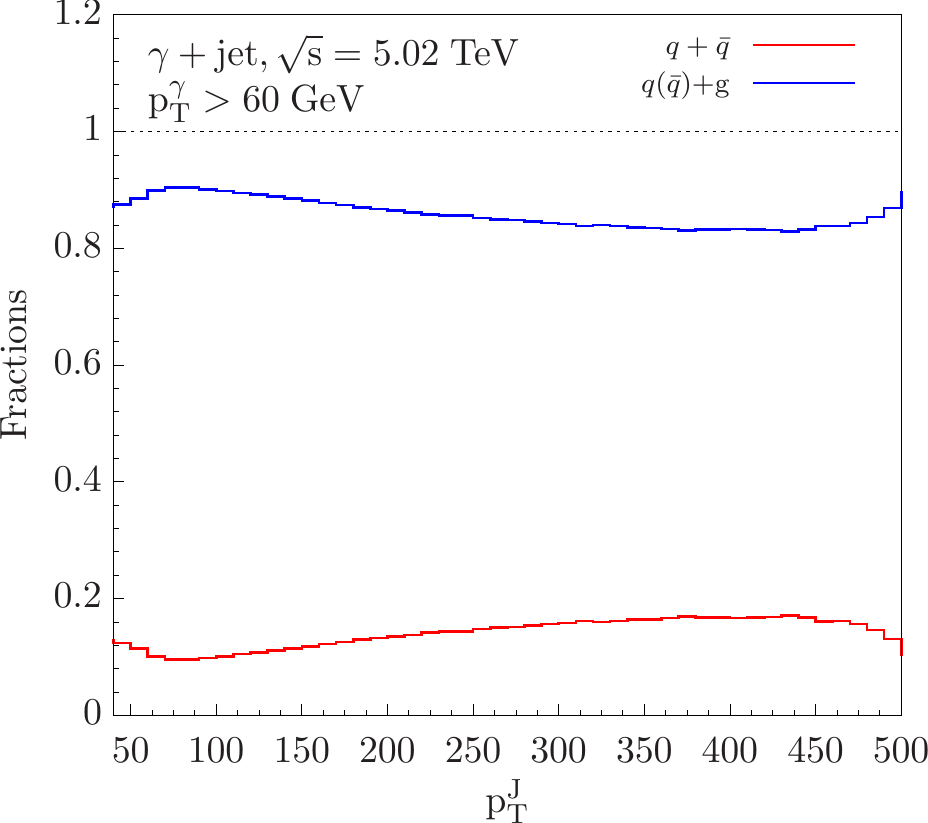} 
\caption{Left panel:  Pythia 8 simulations are compared to CMS measurements of $Z^0$+jet production in p+p collisions at the LHC at $\sqrt{s} = 7$ TeV. Results are shown differentially  versus the jet $p_T^{J}$.  Right panel: the fractional contributions of different subprocesses to  isolated-$\gamma$+jet  production cross sections in p+p collisions at $\sqrt{s}=5.02$.  }
\label{fig-ppbaseline}       
\end{figure}

Before we proceed to heavy ion collisions, it is important to validate  the perturbative calculations in  the simpler proton-proton reactions.
We use Pythia 8~\cite{Sjostrand:2007gs}, which is a popular high energy phenomenology event generator. It utilizes  leading-order perturbative 
QCD matrix elements+parton shower, combined with the Lund string model for hadronization an can can describe well the main properties 
of a p+p event structure. This is shown in the left panel of figure~\ref{fig-ppbaseline}  in comparison to $Z^0$+jet measurements made by the CMS collaboration 
at $\sqrt{s} = 7$ TeV~\cite{Khachatryan:2014zya}.  We also  find good agreement with $\gamma$-tagged jet measurements, see~\cite{Kang:2017frl}.

At leading order two dominant channel for V+jet  are implemented in Pythia, namely $q+\bar q \to V+ g$ and $q(\bar q)+g \to V+q(\bar q)$. We have checked that the $g+g\to V+g$ channel contribution to the cross section is negligible. The utility of vector boson tagging  beyond constraining the recoil jet energy is that it can be used for
flavor selection. We have shown this in the right panel of figure~\ref{fig-ppbaseline} on the example of photon-tagged jets at $\sqrt{s} = 5.1$~TeV. Results indicate that
$\gamma$ tagging or $Z^0$ tagging preferentially selects quark jets and at high $p_T$ the sample has $\sim 80\%$ purity. The accumulated  statistics allows us
to use the fully differential  $d\sigma/dp_{T}^V dp_{T}^J $ to high transverse momenta of the jet and the vector boson.

\subsection{Heavy ion collisions collisions}
\label{subsec-hi}

\begin{figure}[h]
\centering
\includegraphics[width=6.5cm,clip]{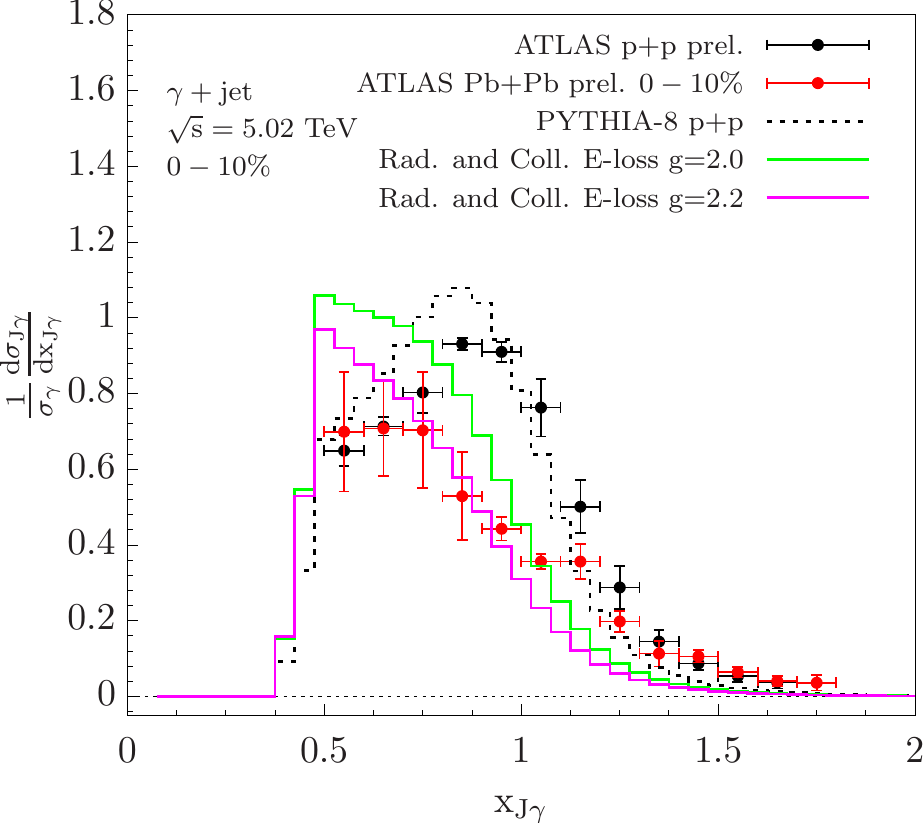} \hspace*{.5cm}
\includegraphics[width=6.5cm,clip]{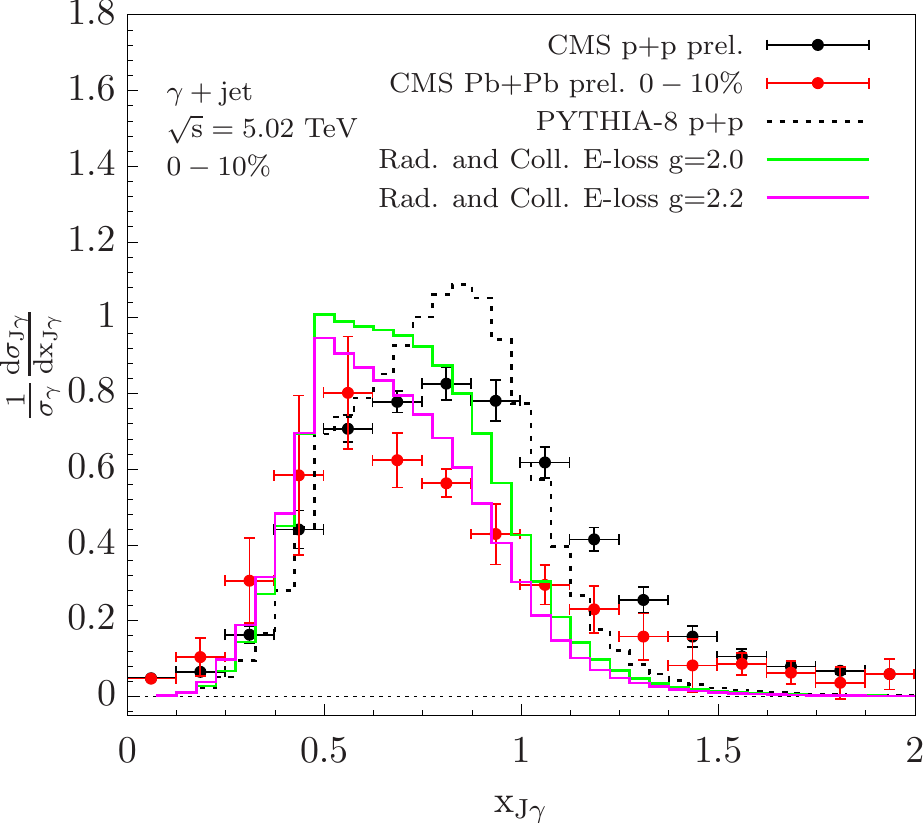} 
\caption{Left panel: The isolated photon-tagged jet asymmetry distributions are shown and compared to ATLAS data in central  
Pb+Pb collisions at the LHC~\cite{ATLAS:2016tor}. The transverse momenta for the isolated photon and the jet are 
$p_T^{\gamma}>60$~GeV and $p_T^J>30$~GeV, respectively. The jet radius parameter is $R=0.4$.  
 Right panel:  same comparison but for CMS data~\cite{CMS:2016ynj} with jet radius $R=0.3$. }
\label{fig-gammajet}       
\end{figure}

We start from the baseline p+p cross sections obtained from Pythia. However, 
in heavy ion collisions we take into account the reaction geometry, the hydrodynamic expansion of the QGP and radiative and collisional energy losses. 
At an impact parameter $|{\bf b}_\perp |$ we evaluate the  cross sections as 
\begin{eqnarray}
\frac{d\sigma^{AA}(|{\bf b}_\perp |) }{dp_{T}^{V}dp_{T}^{J}} 
&=&  \int d^2 {\bf s_\perp}   T_A\left( {\bf s}_\perp - \frac{{\bf b}_\perp}{2}  \right)
T_A\left( {\bf s}_\perp + \frac{{\bf b}_\perp}{2}  \right)  \sum_{q,g} \int_0^1 d\epsilon \frac{P_{q,g}(\epsilon; s_\perp, |{\bf b}_\perp | )}
 {1-f_{q,g}^{\rm loss}(R;s_\perp, |{\bf b}_\perp | )\, \epsilon}
\nonumber \\ 
 &&   \times \,
\frac{ d\sigma^{NN}_{q,g}  \left( p_{T}^{V}, p_{T}^{J}/ \{1-f_{q,g}^{\rm loss}(R; s_\perp, |{\bf b}_\perp | )\,  \epsilon \} \right) }
{dp_{T}^{J}dp_{T}^{V}} .
\label{eq:modify}
\end{eqnarray}
Here hard production follows the binary collision density
  $   \propto T_A\left( {\bf s}_\perp - {{\bf b}_\perp}/{2}  \right) T_A\left( {\bf s}_\perp + {{\bf b}_\perp}/{2}  \right)  $ 
and  we use an optical Glauber model with inelastic nucleon-nucleon scattering cross sections  $\sigma_{\rm in} = 70 $~mb at $\sqrt{s_{NN}} = 5.02$~TeV. 
In equation~\ref{eq:modify} $P_{q,g}(\epsilon) $  is the probability that a fraction $\epsilon $ of the parent parton is redistributed through medium-induced 
multiple gluon emission. We employ the soft gluon emission limit of the of the SCET$_{\rm G}$~\cite{Idilbi:2008vm,Ovanesyan:2011xy} splitting 
functions~\cite{Ovanesyan:2011kn,Fickinger:2013xwa} used to predict inclusive light hadron
attenuation in heavy ion collisions~\cite{Kang:2014xsa,Chien:2015vja}.    We note that  the spectrum coincides with the 
original energy loss derivation of~\cite{Gyulassy:2000fs}. 
Of course, only the fraction of the energy that is redistributed outside of the jet cone leads to jet cross section suppression. 
We evaluate this  fraction as follows
\begin{equation}
f_{q,g}^{\rm loss}(R;{\rm rad+coll}) = 1-  \left(  \int_0^{R }  dr  \int_{\omega_{\rm min}}^E  d\omega \,     \frac{dN^g_{q,g}(\omega,r)}{d\omega dr}    \right)  \Bigg/
 \left(  \int_0^{R_{\rm max} }   dr  \int_0^E  d\omega \,     \frac{dN^g_{q,g}(\omega,r)}{d\omega dr}   \right) .
\label{fradcol}
\end{equation} 
Equation~\ref{fradcol} treats radiative and collisional energy losses on the same footing since for convenience
 we have express this total collisional energy loss as an integral over the spectrum of the medium induced gluons   
\begin{equation}
\Delta E^{\rm coll}_{q,g}({\rm tot.})   = \sum_{i=1}^{N^{\rm tot. \; partons}_{q,g} \ }  \int_{z_i}^\infty   \frac{d \Delta E_i^{\rm coll}}{ d \Delta z}   d   \Delta z 
\;,  \quad     \Delta E^{\rm coll}_{q,g}({\rm tot.})   = \int_0^{\omega_{\rm min}} d\omega  \int_0^{R_{\rm max} }  dr \;     \omega  \frac{dN^g_{q,g}(\omega,r)}{d\omega dr}  .
\label{collshower}
\end{equation}

To lowest leading order in perturbative QCD,   the transverse momentum of the vector boson is balanced by the transverse momentum of the jet,
$p_{T}^V = p_{T}^J$. Higher order processes and  the development of parton showers  introduce deviations from this equality. The exact differential distribution of 
$d\sigma / d p_{T}^V dp_{T}^J$ is also affected  by the jet reconstruction algorithm, jet radius  choice, experimental cuts,  
and detector resolution.   Nevertheless, the quenching and shift of this distribution to smaller values of 
$p_{T}^J$   are currently the  best proxies for jet energy loss. To study these effects one introduces the transverse momentum 
imbalance $\rm x_{JV} = p_{T}^J/p_T^V$. distribution can be obtained from the double differential distribution of V+jet cross section 
\begin{equation}
\frac{d\sigma}{d{\rm x_{JV}}} = \int_{p_{T}^{J, \rm min}}^{p_{T}^{J, \rm max}} d p_{T}^J
\frac{p_T^J}{\rm{x}_{JV}^2}   \frac{d\sigma(p_T^V=p_T^J/{\rm x_{JV}}, p_T^J)}{dp_T^V dp_T^J} \;,
\label{sigAA}
\end{equation}
where $p_{T}^{J, \rm min}$ and $p_{T}^{J, \rm max}$ are matched to the desired cuts of the  experimental measurements.

In figure~\ref{fig-gammajet} we plot the normalized momentum imbalance distributions for the $\gamma$+jet final state in p+p and Pb+Pb collisions at the LHC.  Comparison of the calculations to the  ATLAS measurements~\cite{ATLAS:2016tor} are shown in the left panel.   
 Comparison of the calculations to the  CMS measurements ~\cite{CMS:2016ynj}  are shown in the right panel.  
The black dashed histogram shows  Pythia 8 simulations for the  the $\rm x_{JZ}$ distribution in  p+p collisions.  
The black solid points represent the ATLAS and CMS results in the elementary nucleon-nucleon collisions.
It  can be seen that the $\rm x_{JZ}$ distribution from Pythia 8 simulation is narrower than the one measured by the experiments. 
Part of the reason is that experimental measurements are not unfolded for detector resolution effects. This can be seen on
the example of $Z^0$+jet momentum imbalance by CMS  in Ref. 
for the p+p reference~\cite{Sirunyan:2017jic}. We applied the same smearing functions that  experiment applies to Monte Carlo simulations,  to our 
differential $p_T$ distributions for p+p and Pb+Pb collisions. We got broader ${\rm x_{JZ}}$ distributions and better agreement between the curves to the data points in both p+p and Pb+Pb collisions.
In figure~\ref{fig-gammajet}  we also give results of our theoretical calculations in Pb+Pb collisions. The green and magenta histograms correspond to jet-medium coupling strengths $g=2.0$ and $g=2.2$, respectively,  values that have worked well in describing the light inclusive hadron \cite{Chien:2015vja,Kang:2014xsa} and jet suppression data~\cite{Kang:2017frl} at the LHC. Our results include  both medium-induced radiative energy loss and the parton showers energy dissipation of  through collisional processes in  the QGP. We
see the downshift of ${\rm x_{JV}}$, which  quantitatively agrees with measurements in terms of the difference between p+p and Pb+Pb collisions. 

\begin{figure}[h]
\centering
\includegraphics[width=13.5cm,clip]{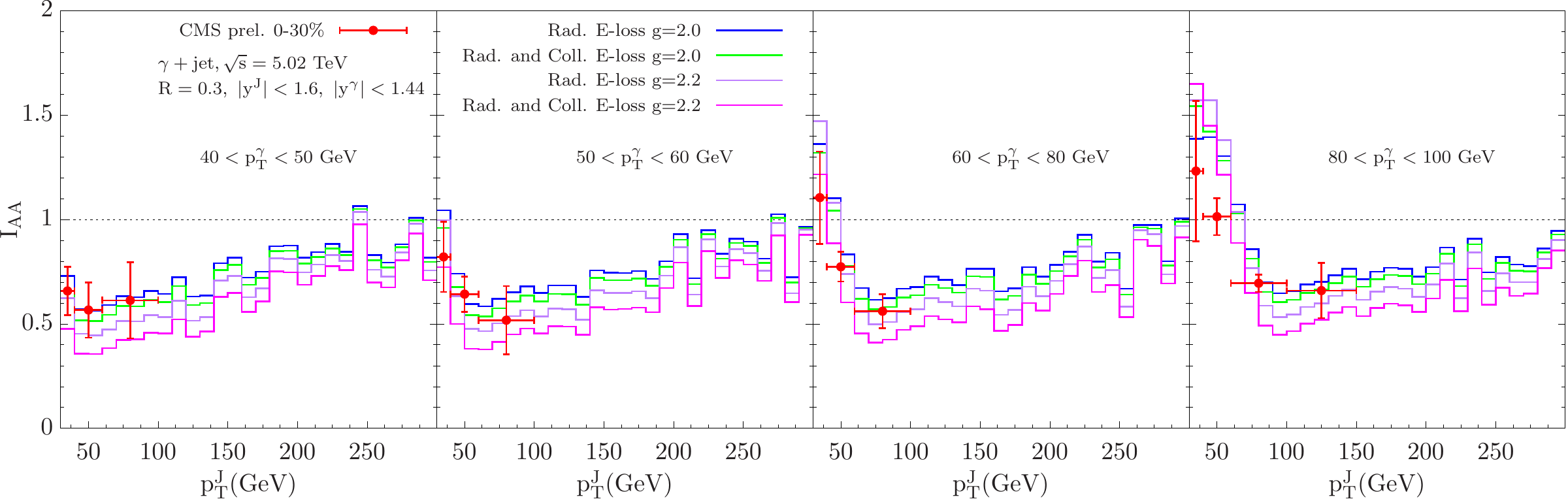}
\caption{Theoretical results for ${\rm I_{AA}}$ compared to CMS data with several different  transverse momentum cuts.  Shown are four different scenarios for energy loss effects with and without collisional energy loss and with couplings between the jet and the medium $g=2.0$ and $g=2.2$}
\label{fig-IAA}       
\end{figure}

Another often studied  observable sensitive to nuclear modification effects in V+jet systems is ${\rm I_{AA}}$.  It  is defined as ratio of the  differential cross section for tagged jets in A+A collisions to the binary collision scaled p+p result. We present  ${\rm I_{AA}}$ in isolated $\gamma$+jet production at the LHC in $0-30\%$ Pb+Pb collisions in figure~\ref{fig-IAA}. By comparing to CMS experimental data, we find that our theoretical results agree with data for a wide kinematic range. In each $p_T^{\gamma}$ window, the jet energy loss effects are shown in four cases and coded with different colors -  two different jet-medium coupling strengths and simulations where collisional energy loss effects are either included or excluded.  Figure~\ref{fig-IAA} illustrates that there is kinematic sensitivity of ${\rm I_{AA}}$ and the largest suppression is observed along the diagonal region of the transverse momenta of the trigger $\gamma$ and the recoil jet. The reason for this behavior in 
 $p_T^{\gamma} \approx p_T^{J}$  arises from the steeper falling cross section in that region. The cross section is suppressed in the region $p_T^J>p_T^{\gamma}$ and  is. enhanced for $p_T^J<p_T^{\gamma}$. This is characteristic of in-medium tagged-jets as discussed in previous works~\cite{Neufeld:2012df,Dai:2012am}.

\begin{table}
\caption{Theoretical predictions for the difference of the average ${\rm x_{JZ}}$ between p+p and Pb+Pb  $0-30\%$ central collisions. The center of mass energy is chosen to be $\sqrt{s} = 5.02$ TeV, the transverse momentum cut for the recoil jet is $p_T^J>30$ GeV, and data is from CMS~\cite{Sirunyan:2017jic}.  }
\label{tab-shifts}       
\centering
\begin{tabular}{l  | c | c | c | c }
\hline
\hline
&\multicolumn{4}{ c }{$\Delta\langle{\rm x_{JZ}} \rangle $}  \\
\hline
$~p_T^Z$ (GeV)& ~$40 - 50$ ~ & ~ $50 - 60$~
  & ~ $60 - 80$ ~ & ~ $80 - 120$ ~ \\ 
 \hline
CMS \cite{Sirunyan:2017jic} & 0.061$\pm$0.059 & 0.123$\pm$0.051  & 0.124$\pm$0.052 & 0.068$\pm$0.042 \\
Rad. + Coll. $g = 2.0$ & 0.022 & 0.050 & 0.075 & 0.086 \\
Rad. + Coll. $g = 2.2$ & 0.024 & 0.058 & 0.093 & 0.119 \\
\hline
\hline
\end{tabular}
\vspace*{0cm}  
\end{table}

To better quantify the shift toward smaller values of the ${\rm x_{JV}}$ distribution, we introduce the mean value of ${\rm x_{JV}}$  and the difference
between proton and heavy ion collisions
\begin{equation}
\langle {\rm x_{JV}} \rangle = \left.\left(\int d{\rm x_{JV}} {\rm x_{JV}} \frac{d\sigma}{d{\rm x_{JV}}}\right)\right/\left(\int d{\rm x_{JV}} \frac{d\sigma}{d{\rm x_{JV}}}\right), 
\quad \Delta\langle {\rm x_{JV}}\rangle = \langle {\rm x_{JV}}\rangle_{\rm pp} -\langle {\rm x_{JV}}\rangle_{\rm PbPb}. 
\end{equation}
The calculated  downshifts of the ${\rm x_{JV}}$ distribution is consistent with the experimental data ~\cite{Sirunyan:2017jic} within the measurement uncertainties for different $p_T^Z$ cuts. More  importantly, when the downshift  is recast into energy loss of the jet recoiling against the  $Z^0$ boson, we find that this experimentally determined jet energy loss corresponds within 15\% accuracy to the theoretically computed energy redistributed out of the jet cone due to in-medium interactions of the
parton shower

\section{Jet substructure}
\label{sec-structure}
We now turn to the question of jet substructure in heavy ion collisions.
\subsection{Momentum sharing distributions}
\label{subsec-sharing}

A new jet substructure observable, called the groomed momentum sharing, was recently proposed and studied using the soft drop jet grooming procedure~\cite{Larkoski:2014wba,Dasgupta:2013ihk}.  It is sensitive to the hard branching in the jet formation and is controlled by the leading-order Altarelli-Parisi splitting functions.  Having  reconstructed a jet of radius $R$  using the anti-$k_T$ algorithm,  one can recluster the jet using the Cambridge/Aachen algorithm and goe through the branching history, dropping the soft branch until  $ z_{cut} < {\min(p_{T_1},p_{T_2})}/{(p_{T_1}+p_{T_2)}} \equiv z_g$.  One can also demand that the angular separation between the two branches is greater than an angular scale $\Delta$,  $ \Delta < \Delta R_{12} \equiv r_g$.   By selecting the  groomed jet radius $r_g$, one gains access to the momentum sharing distribution $p(z_g)$ at different splitting angles,  and can study the angular  distribution $\rho{z_g}$.
We evaluate
\begin{equation}
    p_i(z_g)=\frac{\int_{k_\Delta}^{k_R}dk_\perp\overline{\cal P}_i(z_g,k_\perp)}{\int_{z_{cut}}^{1/2}dx\int_{k_\Delta}^{k_R}dk_\perp\overline{\cal P}_i(x,k_\perp)}
\quad {\rm where} \; \;  k_\Delta = \omega x(1-x)\tan{\frac{\Delta}{2}},\;  k_R = \omega x(1-x)\tan{\frac{R}{2}}.  \label{pzg}
\end{equation}
The symmetrized splitting functions are  $\overline{\cal P}_i(x,k_\perp) = 
\sum_{j,l}\Big[{\cal P}_{i\rightarrow j,l}(x,k_\perp)+{\cal P}_{i\rightarrow j,l}(1-x,k_\perp)\Big]$ and in the presence of a QGP,
$    {\cal P}_{i\rightarrow jl}(x, k_\perp)={\cal P}^{vac}_{i\rightarrow jl}(x, k_\perp)+{\cal P}^{med}_{i\rightarrow jl}(x, k_\perp)$. 
The medium-induced component were calculated using soft-collinear effective theory  with Glauber gluon interactions ($\rm SCET_G$)~\cite{Idilbi:2008vm, Ovanesyan:2011xy,Ovanesyan:2011kn,Fickinger:2013xwa}.

\begin{figure}[h]
\centering
\includegraphics[width=7.cm,clip]{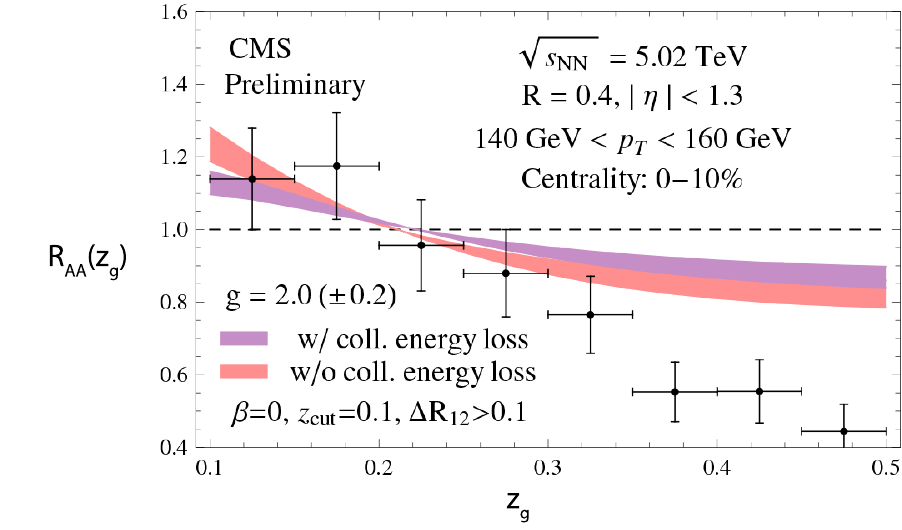} \hspace*{-.1cm}
\includegraphics[width=7.cm,clip]{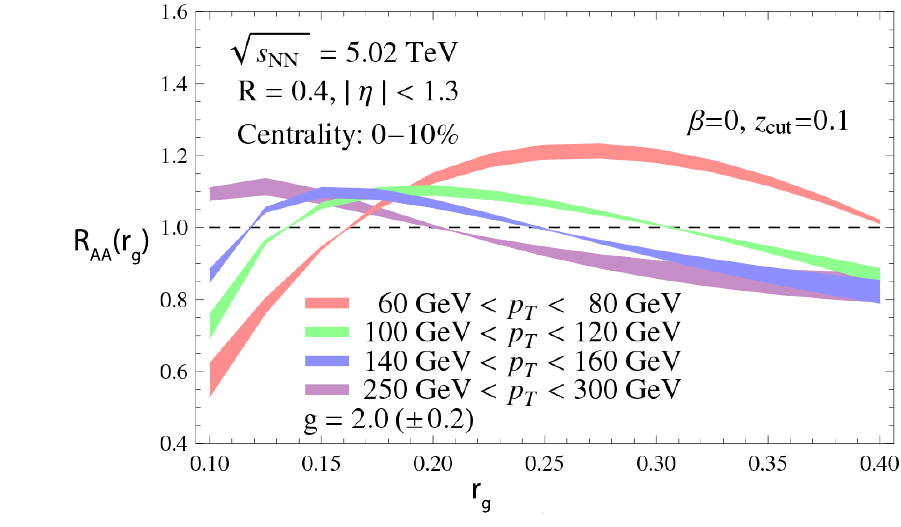} 
\caption{Left panel: theoretical calculations compared to preliminary CMS data for the ratio of momentum sharing distributions of inclusive anti-$k_T$ $R=0.4$ jets in  central Pb+Pb and p+p collisions at $\sqrt {s_{\rm NN}} = 5.02$ TeV. Purple and red bands correspond to calculations with and without the implementation of collisional energy loss. Right panel:  the groomed jet radius modification of inclusive jets in proton-proton and central lead-lead collisions at $\sqrt {s_{\rm NN}} = 5.02$ TeV. We show results for four $p_T$ bins with $60~{\rm GeV}< p_T < 80~{\rm GeV}$ , $100~{\rm GeV}< p_T < 120~{\rm GeV}$, $140~{\rm GeV}< p_T < 160~{\rm GeV}$  and $250~{\rm GeV}< p_T < 300~{\rm GeV}$.}
\label{fig-splitting}       
\end{figure}

 Analytic and  numerical considerations show that for $x < 1/2$ in Eq.~(\ref{pzg}) medium-induced component goes as $1/x^2$ ~\cite{Ovanesyan:2011kn}.
We thus expect that the momentum sharing distribution will be enhanced at small $z_g$ and
suppressed near $z_g=1/2$. Numerical results~\cite{Chien:2016led} are shown in the left panel of figure~\ref{fig-splitting} and compared to preliminary CMS data.
The parameters for the groomed  soft-dropped jets are $\beta = 0$, $z_{cut} = 0.1$ and $\Delta R_{12} > 0.1$. We estimate the theoretical uncertainty by varying the jet-medium coupling $g=2.0 \pm 0.2$. Analogous calculations in the soft gluon emission limit have also  been presented \cite{Mehtar-Tani:2016aco,Chang:2017gkt}. The finalized CMS data \cite{Sirunyan:2017bsd} changed significantly and is now in excellent agreement with the theoretical predictions of~\cite{Chien:2016led}. This signifies  the  need for reliable pQCD theory to interpret heavy ion data and motivates further studies, such as the  groomed
jet radius distribution. The anticipated modification is shown in the    right panel of figure~\ref{fig-splitting}  for several different $p_T$ bins. The peak of this distribution reflects the medium enhancement of large-angle splitting for hard branching processes.

\subsection{Jet shapes}
\label{subsec-shapes}

A complementary observable, in fact the first one that has been quantitatively studied in relation to inclusive jet modification in heavy ion collisions~\cite{Vitev:2008rz}, is the jet shape. The jet shape~\cite{Ellis:1992qq} describes the transverse energy profile inside a jet of size $R$, reconstructed using a jet algorithm. 
The integral jet shape, defined as the fraction of the transverse energy $E_T$ of the jet within a subcone of size $r$,  reads
\begin{equation}  
  \Psi_J(r)=\frac{\sum_{i,~d_{i\hat n}<r} E^i_T}{\sum_{i,~d_{i\hat n}<R} E^i_T}, \quad {\rm where } 
 \; \;  d_{i\hat n}=\sqrt{(\eta_i-\eta_{jet})^2+(\phi_i-\phi_{jet})^2}
\end{equation}
In heavy ion collisions, the differential jet shape $\rho(r)=\frac{d}{dr}\Psi(r)$ is more sensitive to in-medium modification, which can be presented 
as $M_{\rho}(r) =  { \rho^{AA}(r)}/{ \rho^{pp}(r) }$.  

\begin{figure}[h]
\centering
\includegraphics[width=7.cm,clip]{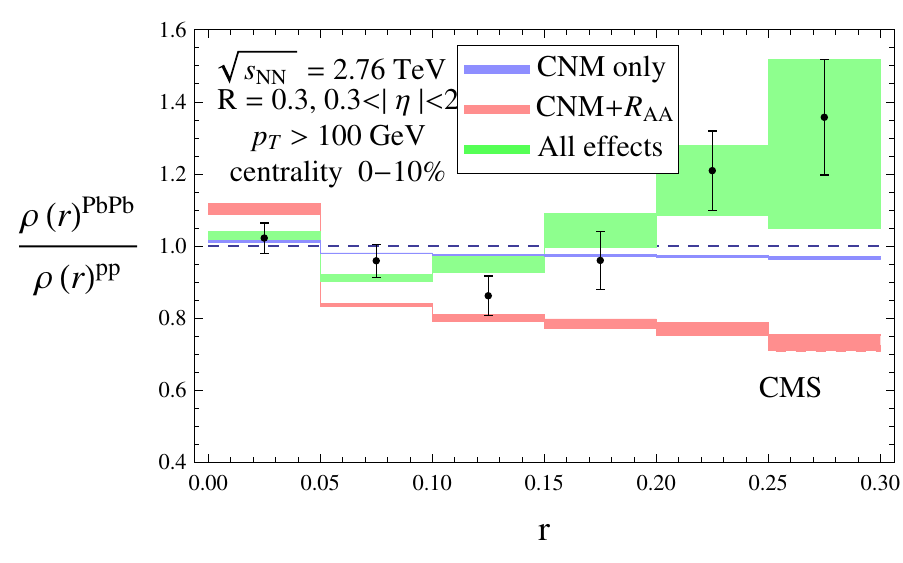} \hspace*{-.1cm}
\includegraphics[width=7.cm,clip]{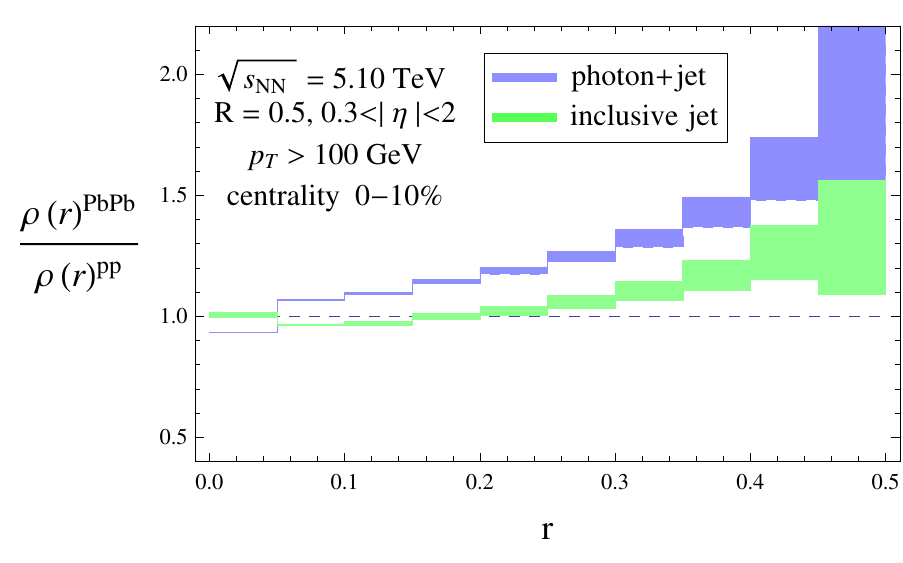} 
\caption{Left panel:  the modification of differential jet shapes of inclusive jets in Pb+Pb central collisions at  $\sqrt{s_{\rm NN}}=2.76$~TeV at the LHC Different
contributions are shown and compared to CMS data.  Right panel:  theoretical predictions for the differential jet shape  modification of inclusive and 
photon-tagged jets with $R=0.3$ in  Pb+Pb collisions at $\sqrt{s_{\rm NN}}\approx 5.1$~TeV at the LHC.}
\label{fig-shapes}       
\end{figure}

A robust theory of jets in high energy nuclear collisions should describe, within its regions of applicability, the modification of the jet shapes in addition to the
attenuation of the inclusive and dijet cross sections~\cite{He:2011pd}.  We have taken the opportunity to address this key issue in~\cite{Chien:2015hda},
building upon an improved description of   $\rho^{pp}(r)$~\cite{Chien:2014nsa}  in the framework of SCET~\cite{Bauer:2000yr}.
Our results are shown in figure~\ref{fig-shapes}, where we impose cuts on the jet transverse momentum $p_T > 100$~GeV and pseudo-rapidity 
$0.3<|\eta|<2.0$ of jets. We take the coupling between the jet and the QGP to be $g=2$ and the theoretical uncertainty is estimated by varying the jet energy scales $\frac{1}{2}\mu_{j_R}<\mu<2\mu_{j_R}$ in the calculations.  Several contributions to the jet shape modification are shown. The blue band corresponds to the initial state cold nuclear 
matter effects. The effect of quenching is added in the  red band, which enhances the fraction of quark jets that are narrower than the gluon jets. 
The green band correspond to the full calculation the green band is the full calculation and adds the event-by-event broadening.   It is evident that the 
jet shape modification is highly nontrivial and we find that it compares well to the  data from CMS~\cite{Chatrchyan:2013kwa} with jet radius $R=0.3$.

We also present  predictions for the modification of jet shapes in Pb+Pb collisions at $\sqrt{s_{\rm NN}}\approx 5.1$~TeV, focusing on the  differences
between inclusive jets and tagged jets. The results are given in the right panel of figure~\ref{fig-shapes}. While  $M_{\rho}(r)$ for jet shapes at $\sqrt{s_{\rm NN}}\approx 5.10$~TeV is similar to the one at $\sqrt{s_{\rm NN}}= 2.76$~TeV, the modification of $\gamma$-tagged jet shapes is more pronounced.  Photon-tagged jet shapes also exhibit
a different modification pattern. Since photon tagging selects primarily quark jets, see previous section, the quenching does not play a significant role and the
enhancement toward the periphery of jets is more pronounced. Other examples of jet shape calculations include~\cite{Brewer:2017fqy,Tachibana:2017syd}. 

\section{Conclusions}
\label{sec-concl}

To summarize,  in these proceedings we reported  results  on vector boson-tagged jet production~\cite{Kang:2017xnc}
and jet substructure~\cite{Chien:2015hda,Chien:2016led}  in heavy ion collisions.   
We presented a new study of  $\gamma$-tagged  or $Z^0$-tagged  jets in Pb+Pb 
reactions at the LHC a center-of-mass energy per nucleon pair of 5.02~TeV. Good description of the mean 
momentum imbalance shift  $ \Delta \langle {\rm x_{JV}} \rangle$  and the tagged jet nuclear modification 
factor ${\rm I_{AA}}$ can be achieved in a framework that accounts for the QGP-induced radiative and  collisional 
dissipation of the parton shower energy outside of the jet cone. At the heart of such quenching effects for 
reconstructed jets is the in-medium parton branching -- qualitatively and quantitatively different from the standard
Altarelli-Parisi splittings in the vacuum. New techniques, such as the study of the subjet momentum 
sharing distributions,  have recently emerged and can probe precisely this parton dynamics.   
We also presented the first calculation of this subjet momentum 
sharing distribution in heavy ion collisions. The validation of the theoretical predictions by  recent experimental measurements
complements our earlier results on the modification of the jet shape  and  corroborates the  perturbative picture of
parton shower modification in the medium. 
Finally, we found that the modification of $\gamma$-tagged and $Z^0$-tagged jet substructure can differ substantially
from the corresponding  substructure modification of inclusive jets.   Future experimental results for such observables can 
significantly enhance our understanding of in-medium QCD dynamics.

%

\end{document}